\documentclass [12pt] {iopart}
\usepackage {graphicx}
\begin {document}

\title [A theory of MHD instability of an inhomogeneous plasma jet]
{A theory  of MHD instability of an inhomogeneous plasma jet}

\author {A. S. Leonovich}

\address {Institute of Solar-Terrestrial Physics SB RAS, Irkutsk, Russia}
\ead {leon@iszf.irk.ru}
\date{\today}

\begin {abstract}
\begin {sloppypar}
A problem of the instability of an inhomogeneous axisymmetric plasma jet in a
parallel magnetic field is solved. The jet boundary becomes, under certain
conditions, unstable relative to magnetosonic oscillations (Kelvin-Helmholtz
instability) in the presence of a shear flow at the jet boundary. Because of
its internal inhomogeneity the plasma jet has resonance surfaces, where
conversion takes place between various modes of plasma MHD oscillations.
Propagating
in inhomogeneous plasma, fast magnetosonic waves drive
the Alfven and slow magnetosonic oscillations,
tightly localized across the magnetic shells, on the
resonance surfaces. MHD oscillation energy is absorbed
in the neighbourhood of these resonance surfaces. The resonance surfaces
disappear for the eigen-modes of slow magnetosonic waves propagating in the
jet waveguide. The stability of the plasma MHD flow is determined by competition
between
the mechanisms of shear flow instability on the boundary and
wave energy dissipation because of resonant MHD-mode coupling. The problem is solved
analytically, in the WKB approximation, for the plasma
jet with a boundary in the form of a tangential discontinuity
over the radial coordinate.
The Kelvin-Helmholtz instability develops if plasma flow velocity in the jet
exceeds the maximum Alfven speed at the boundary.
The stability of the plasma jet with a smooth boundary layer is investigated
numerically for the basic modes of MHD oscillations, to which the
WKB approximation is inapplicable. A new "global" unstable mode of
MHD oscillations has been discovered which, unlike the Kelvin-Helmholtz
instability, exists for any, however weak, plasma flow velocities.

\end {sloppypar}
\end {abstract}


\section {Introduction}

Shear plasma flows in magnetic field are encountered in many problems in
magnetohydrodynamics. Of most interest are usually unstable
oscillations developing in a shift layer.  Thus, many kinds of geomagnetic field
oscillations related to the Kelvin-Helmholtz instability develop on
the Earth's magnetospheric boundary when the solar wind plasma flows around it
\cite{McKenzie 1970a, Kivelson Pu 1984}. Similar instabilities arise in
differentially rotating plasma shells of stars \cite{Watson1981}. The
problem of the instability of the plasma configuration boundary in experimental
installations where the plasma was confined magnetically
has been discussed widely enough \cite{Rosenbluth1957, Lukiyanov1975}. In such installations,
plasma injected along the magnetic field lines becomes unstable \cite{Perkins1963}.

Analytical studies devoted to shear flow stability are often
stated for two-layer medium models where fluid, gas or plasma
move in two homogeneous half-spaces separated by a flat shift layer of
velocity \cite{McKenzie 1970a, Landau 1944}. Using such models enables one
to progress far enough in constructing analytical solutions to
hydrodynamic (or magnetohydrodynamic) equations describing unstable
oscillation modes. However, real shear flows occur, as a rule, in an
inhomogeneous medium in a layer of finite thickness. Constructing analytical
solutions in such models is only possible for several extreme cases
\cite{Thorpe 1969}. Solutions to corresponding equations are often obtained
by numerical integration \cite{Miura 1992}. As a rule, all effects
related to medium inhomogeneity are ascribed, in these models, to the
shift layer, while the medium away from it is supposed to be homogeneous. The
solutions obtained in this way have a rather strong limitation regarding
their applicability area as well.

In many real events the medium remains inhomogeneous (though the scale of the
inhomogeneity is smaller than in the shift layer) even far from the shift
layer. This inhomogeneity can also play a considerable role in forming
the conditions under which the unstable oscillation modes develop in the
shift layer \cite{Glassmeier 1996}. For example, in the inhomogeneous layer, the
resonance surfaces can exist where there is a coupling of various modes of
MHD oscillations. When fast magnetosonic (FMS) waves propagate in
inhomogeneous plasma they can drive the Alfven and slow magnetosonic (SMS)
oscillations tightly localized across magnetic shells, on the resonance
surfaces \cite{LeonKoz 2009}. This results in the oscillation energy
absorbed by particles of the background plasma, heating up in the process
\cite{ChenHas 1974, Erdelyi2004}. The mechanism stabilizes the unstable modes.

Shear flows bounded in space also have their specific features. Such jet
flows arise, for example, when plasma clusters are injected into magnetic
traps along the magnetic field lines \cite{Azovsky 1967}. The same
conditions arise when plasma filaments erupt from the Sun surface in the
corona \cite{Filippov 2009}, and also when the solar wind plasma flows round
a planet's magnetosphere \cite{McKenzie 1970a, McKenzie 1970b}. The model of a cylindrical plasma
jet propagating parallel to the magnetic field lines is obviously closest
to reality in all these situations. There are a few studies devoted
to the stability of hydrodynamic and magnetohydrodynamic flows in
cylindrical models (see \cite{McKenzie 1970b, Drazin 1966}).

This work tackles the problem of stability of a cylindrical plasma jet in
a parallel magnetic field. The plasma in the jet is assumed to be homogeneous
over the azimuth and inhomogeneous over the radius. The velocity of plasma
in the jet is supposed to be homogeneous. The shear flow occurs in a layer of
finite thickness at the plasma jet boundary. For a qualitative understanding
of the structure and dynamics of the unstable oscillation modes, this
problem is solved in the WKB approximation over the radial coordinate.
The boundary of the plasma jet is assumed to be in the form of a
tangential discontinuity. The solution to the problem when the boundary has
the form of a smooth transition layer is obtained numerically for the first
harmonics of unstable oscillations, for which the WKB approximation is
inapplicable.

This paper is structured as follow. The model of the medium is presented and
the basic equations of the problem under study are derived in Section 2.
Section 3 is a qualitative examination of the structure of the radial
component of the wave vector in the WKB approximation, as well as deriving
the boundary conditions and the matching condition on the jet boundary for
unstable MHD oscillations in question. Section 4 examines the structure of
the forced modes and eigen-modes of MHD oscillations of the plasma jet in the
WKB approximation. The dependence on the plasma flow velocities of the
increment of unstable oscillations of the plasma jet (with a boundary
in the form a tangential discontinuity) is analytically studied in Section 5.
Section 6 provides a numerical solution to the same problem for a plasma jet
with its boundary in the form of a smooth transition layer. Section 7
explores the instability of the "global modes" of plasma jet oscillations.
The Conclusion lists the main results of this work.

\section {Model medium and basic equations}

Let us consider a model cylindrical plasma jet presented in fig.1-2.
Let us introduce a cylindrical coordinates system ($r, \phi, z $) where the
origin $r=0$ coincides with the jet axis. We assume the background
magnetic field to be directed along the $z $ axis and be homogeneous (but
not identical) inside and outside the plasma jet. In calculations for the jet
boundary in the tangential discontinuity approximation, the parameters of the
medium on its conventional boundary $r=r_b $ will have a subscript $ _I $ \--
inside, and $ _ {II} $ outside. We will consider the plasma in the jet to
be moving along the $z $ axis at velocity $v_0$, and the plasma outside
of the jet to be immobile (see fig.1). Transition from the jet parameters
to the parameters outside occurs within a narrow transition layer of thickness $
\Delta_r\ll r_b $. The plasma density distribution over the radius will
be considered as maximum at the jet axis and decreasing to a minimum toward
the boundary. We assume the magnetic field inside the jet to be greater
than outside. The distribution of the Alfven speed $A=B_0/\sqrt
{4\pi\rho_0} $ over the radius has the form presented qualitatively
in fig.1-2. Such a distribution of plasma parameters occurs in
magnetic arches on the Sun, in the magnetotail of the Earth's
magnetosphere, as well as in laboratory installations with magnetic plasma
confinement of the theta-pinch type.

\begin {figure} \label {f1}
{\includegraphics [width=10cm] {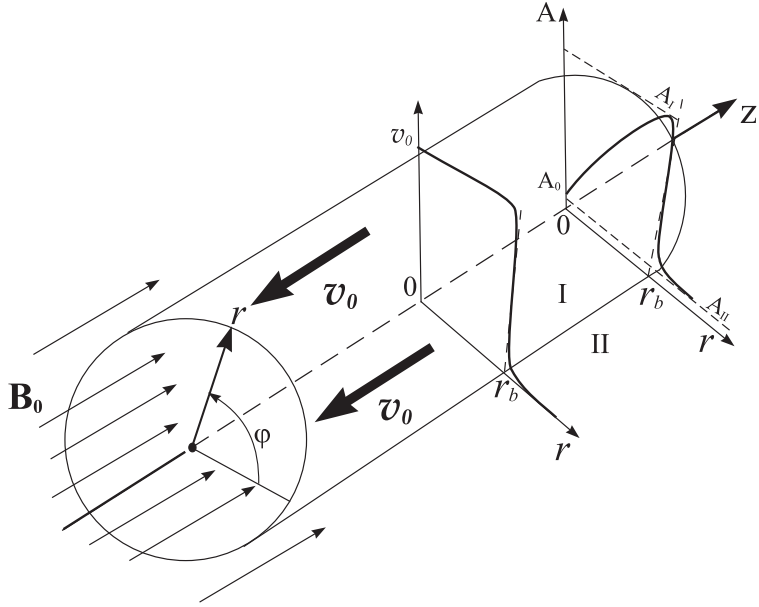}}
  \caption {
A model of cylindrical plasma jet whose plasma motion is directed
against the background magnetic field $ \bi {B_0} $. The distributions of the
velocity profile of moving plasma $ {v_0} (r) $ and of the Alfven speed $A
(r)$ are presented schematically. }
\end {figure}

To describe such a plasma configuration we used a set of ideal MHD
equations of the form
\numparts
\begin {eqnarray}
\label {Eq1_a}
\rho\frac {d\bar {\bi {v}}} {dt} &=&-\-\nabla\bar {P} + \frac1 {4\pi} [\mathop {\mathrm {curl}} \bar {\bi {B}} \times\bar {\bi {B}}], \\\label {Eq1_b}
\frac {\partial\bar {\bi {B}}} {\partial {} t} &=& \mathop {\mathrm {curl}} [\bar {\bi {v}} \times\bar {\bi {B}}], \\\label {Eq1_c}
\frac {\partial\bar {\rho}} {\partial {} t} &+& \nabla (\rho\bar {\bi {v}}) =0, \\\label {Eq1_d}
\frac {d} {dt} \frac {\bar {P}} {\bar {\rho} ^ {\gamma}} &=&0, \end {eqnarray}
\endnumparts
where $ \bar{\bi {B}}, \ \bar{\bi {v}} $ are vectors of the magnetic
field and velocity of the plasma motion, $ \bar{\rho}, \ \bar{P} $ are the
plasma density and pressure, $ \gamma=5/3$ is the adiabatic index. Let us
assume the wave-related disturbance to be weak enough, allowing for
the initial set of equations to be linearized. Let us denote the parameters
of the unperturbed plasma with a subscript of zero, while leaving the
wave-related parameters unindexed: $ \bar{\bi {B}} =
\bi{B_0} + \bi{B}, \ \bar{\bi {v}} = \bi{v_0} + \bi{v}, \ \bar{\rho} = \rho_0
+\rho, \bar{P} =P_0+P $. In the zero approximation the $r $ component of
equation (\ref {Eq1_a}) in steady state ($ \partial/\partial t=0$) yields the
equilibrium condition for a plasma configuration
\begin {equation} \label {Eq_2}
P_0 +\frac {B_0^2} {8\pi} =const,
\end{equation}
which determines an equilibrium distribution of the plasma pressure $P_0 (r)
$ for a fixed distribution of $B_0 (r) $. This pressure determines the
distribution of the sound velocity in plasma $S =\sqrt {\gamma P_0/\rho_0} $
and a corresponding distribution of SMS-wave velocity $C_s=AS/\sqrt {A^2+S^2}
$ in fig.~2. Let us assume the magnetic field to be almost
constant inside and outside the plasma jet, changing only in a thin
transition layer of thickness $ \Delta_r\ll r_b $. Then it follows from the
equilibrium condition (\ref {Eq_2}) that the plasma pressure also
varies inside the transition layer only. We denote the component of the
vector of the disturbed plasma velocity in a wave in the $r $ axis direction
$v_r =\mathrm {d} \zeta/\mathrm {d} t \equiv\partial\zeta/\partial t +
(\mathbf {\bi {v} _0\nabla}) \zeta $, where $ \zeta $ is the displacement
of a plasma element. Let us consider the harmonic of a wave in the form $
\exp (ik_zz+im\phi-i\omega t) $, where $k_z $ is the component of the wave
vector in the $z $ axis direction, $m=0,1,2,3... $ is azimuthal wave number,
$ \omega $ is wave frequency. Linearizing the set of equations (\ref
{Eq1_a})\--(\ref {Eq1_d}) and expressing the other components of the
oscillation field through $ \zeta $, we obtain:
\numparts
\begin {eqnarray}
\label {Eq3_a} v_r =-i\bar {\omega} \zeta, \ \ \ \ \ v_\phi = - \frac {1} {K_s^2} \left (A^2 +\frac {K_A^2S^2} {\chi_S^2} \right) \frac {m} {\bar {\omega} r^2} \frac {\partial r\zeta} {\partial r}, \\
v_z = - \frac {k_zK_A^2S^2} {\bar {\omega} \chi_S^2r} \frac {\partial r\zeta} {\partial r}-v_0 ^\prime\zeta, \nonumber
\\ \label {Eq3_b}
B_r =-ik_zB_0\zeta, \ \ \ \ B_\phi =-\frac {k_zB_0} {\bar {\omega}} v_\phi, \ \ \ B_z =-\frac {K_A^2B_0} {\chi_S^2} \left (1-\frac {k_z^2S^2} {\bar {\omega} ^2} \right) \frac {1} {r} \frac {\partial r\zeta} {\partial r}-B_0 ^\prime\zeta, \nonumber \\
\label {Eq3_c}
P =-\gamma P_0\frac {K_A^2} {\chi_S^2} \frac {1} {r} \frac {\partial r\zeta} {\partial r} + \left (\frac {B_0^2} {8\pi} \right)^\prime\zeta,
\end {eqnarray}
 \endnumparts
where
\begin {eqnarray*}
K_A^2=1-\frac {k_z^2A^2} {\bar {\omega} ^2}, \ \ \ \ \ \ \ K_s^2=K_A^2-\frac {m^2A^2} {r^2\bar {\omega} ^2}, \\ \chi_S^2=1-\frac {m^2/r^2+k_z^2} {\bar {\omega} ^2} \left (A^2+S^2-\frac {k_z^2A^2S^2} {\bar {\omega} ^2} \right), \end {eqnarray*}
$ \bar {\omega} = \omega-k_z v_0$ is an oscillation frequency modified by
Doppler's effect. For the displacement $\zeta $ we obtain the equation
\begin {equation} \label {Eq_4}
   \frac {\partial}{\partial r} \, \, \frac{\rho_0\Omega^2} {k_r^2} \, \, \frac{1}{r} \frac{\partial r\zeta} {\partial r} + \rho_0\Omega^2\zeta=0,
\end {equation}
where $ \Omega^2 =\omega^2-k_z^2A^2$,
\begin {eqnarray}\label{Eq_5}
k_r^2 =\frac{\bar{\omega}^4}{\bar {\omega}^2 (A^2+S^2)-k_z^2A^2S^2}-k_z^2-\frac {m^2}{r^2} = \\
k_z^2\left (\frac {\bar {\omega} _A^4 / (1 +\beta ^ *)} {(\bar {\omega} _A^2-\bar {\omega} _S^2)}-1-\frac {m^2} {k_z^2r^2} \right) = \nonumber \\
\frac{k_z^2}{1 +\beta ^ *} \frac{(\bar {\omega} _A^2-\bar {\omega} _ {A1} ^2) (\bar {\omega} _A^2-\bar {\omega} _ {A2} ^2)} {(\bar {\omega} _A^2-\bar {\omega} _ {s} ^2)}, \nonumber
\end {eqnarray}
and the notations are $ \bar {\omega} _A =\bar
{\omega}/k_zA (r) $, $ \bar {\omega} _S =\sqrt {\beta ^* / (1 +\beta ^ *)} $,
$ \beta ^ * = S^2/A^2$, whereas $ \bar {\omega} _ {A1} ^2, \bar {\omega} _ {A2}
^2$ are the roots of a biquadratic (with respect to $ \bar {\omega} _ {A} $)
equation $k_r^2=0$.

\begin {figure} \label {f2}
{\includegraphics [width=10cm] {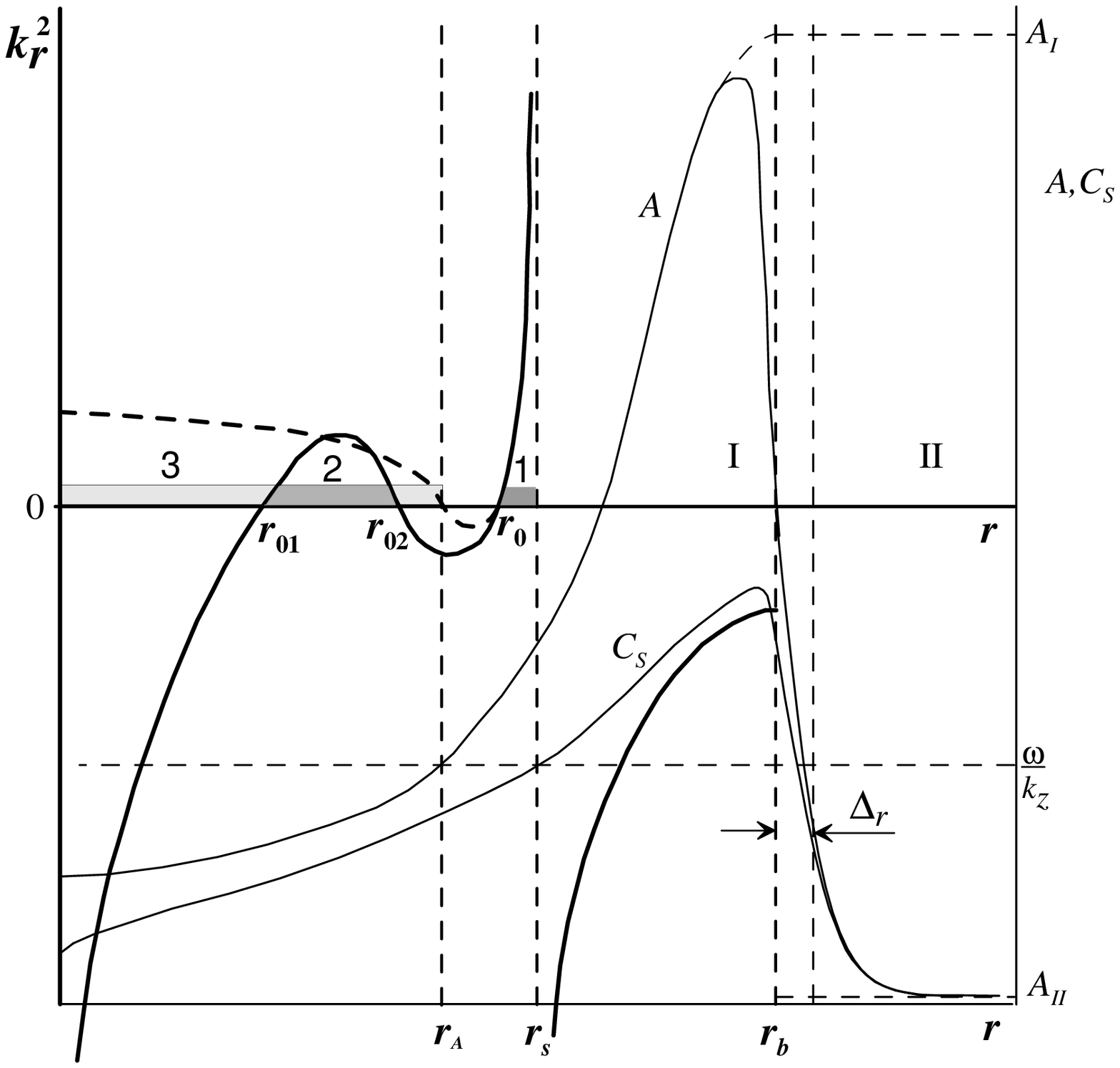}}
  \caption {
Distribution of the Alfven speed $A (r) $ and velocities of the SMS waves
$C_s (r) $ inside and outside of the plasma jet (the thin lines and right-hand
vertical axis). Distribution of the squared WKB component of the wave
vector $k^2_r (r) $ in the plasma jet (the thick lines; for the mode $m=0$
dashed line, left vertical axis). The $r_A $ and $r_S $
coordinates correspond to the resonance surfaces for the Alfven and
SMS-oscillations in the jet, respectively; $r_0, r _ {01}, r _ {02} $ are the
turning points. The numerals and shades of gray demonstrate the transparency
regions: (1) for SMS waves, (2) for FMS waves  ($m\neq 0$), and (3)
$m=0$.
}
\end {figure}

Note that the expression $ \beta ^ * $ coincides, within a factor close to
unity, with the well known parameter $ \beta=8\pi P_0/B_0^2$ - the gas-kinetic
plasma to magnetic pressure ratio. It can be seen from (\ref {Eq_4}) that
$k_r^2$ is the square of the $r $-component of the wave vector in the WKB
approximation when the solution to (\ref {Eq_4}) may be presented in the form
$ \zeta\sim \exp (i\int k_rdr) $.

\section {The distribution of $k_r^2 (r) $, the matching conditions on the
plasma jet boundary and boundary conditions}

Solving the problem in the WKB approximations is determined by the
magnitude of the wave vector component $k_r^2 (r)$ on both sides of the
plasma jet boundary presented in the form of a tangential discontinuity. We
will analyse the behaviour of $k_r^2 (r) $ inside and outside the jet. For
convenience, our subsequent calculations will involve the frame of
reference moving with the flux plasma at velocity $ \bi {v} _0$.
In this frame of reference plasma is immobile in the plasma flux rope, while
moving at velocity $-v_0$ outside of it. The distribution of $k_r^2 (r)
$ in the plasma flux rope is presented qualitatively in fig. 2. This figure
presents the distribution of $k_r^2 (r) $ for such values of $m, k_z $ and $
\omega $ for which all possible resonance surfaces and turning points
are present in the plasma flux rope.

The turning points are determined by zeros of the function $k_r^2 (r) $. In
the distribution in fig.~2 their number can vary from one ($r_0$) to three
($r_0, r _ {01}, r _ {02} $). The number of turning points is determined by
the parameters $m, k_z $ and $ \omega $. Thus, for the axisymmetric mode
$m=0$,
the turning point $r _ {02} $ is absent, whereas another one \- - $r _ {01} $
coincides with the point $r_A $, that determines the location of the
resonance surface for Alfven wave when $m\neq 0$. A transparency region
(where $k_r^2 (r)> 0$) can exist for the FMS waves in the plasma flux rope -
this region is located in the range $r _ {01} \leq r \leq r _ {02} $ when $m\neq 0$,
and in the range $0 \leq r \leq r _ {A} $ when $m=0$. The transparency region
for SMS waves is located in the interval $r _ {0} \leq r \leq r _ {s} $
(where $r_s $ is a resonance surface for the SMS oscillations).

Resonance surfaces are determined by the singular points of equation
(\ref {Eq_4}) where the coefficient of the higher derivative reduces to zero.
One of them \--- the Alfven resonance point $r_A $ determined by the equation
$ \Omega^2 (r_A) =0$ \--- is located in the opacity region in the interval $ (r _
{02}, r _ {0}) $. When $m=0$ the point $r_A $ is the turning point and
is not singular (the coefficient of the higher derivative here does not
reduce to zero). The second singular point \--- magnetosonic resonance point
$r_s $ \--- is determined by the denominator in the expression (\ref {Eq_5})
becoming
zero, yielding a local dispersion equation for SMS waves when $ |k_r^2
|\rightarrow\infty $: $ \omega^2=k_z^2C_s^2 (r_s) $. The point $r_s $ is
located farther along the radius than the turning point $r _ {0} $, and the
transparency region for the SMS waves is located between them. The opacity
region is in the range $r_s <r <r_b $.

It is evident from (\ref {Eq_5}) that the behaviour of $k_r^2 (r) $ in the
range $0 <r <r_b $ depends on the magnitude of $ \bar {\omega} _A (r) $ at
the ends of the interval. It is possible to see the distribution of $k_r^2 (r) $
mentally moving the function $k_r^2 (r) $ shown in fig.2 from left to right.
When $ \omega^2_0 <\omega _ {SI} ^2$ (where $ \omega^2_0\equiv \bar {\omega}
_A^2 (0) $, $ \omega _ {SI} ^2=const $ is the magnitude of $ \bar {\omega}
_S^2$ in a plasma flux rope) we have $k_r^2 (r) <0$ over the entire
cross-section of
the jet, i.e. the entire jet is an opacity region. It is possible to
regard the entire interval $0 <r <r_b $ as the part of the $k_r^2 (r) $ plot,
presented in fig.2, corresponding to the opacity region $r_s <r <r_b $.
The point of magnetosonic resonance $r_s $ is absent from the system
\--- the rest of the plot in fig. 2 can be imagined on
the left of the point $r=0$. When $ \omega^2_0$ increases (due to growing
parallel phase velocity $ \omega/k_z $) this plot moves from left
to right in the range $0 <r <r_b $. The resonance surface for
SMS waves $r_s $ (for $ \omega _ {SI} ^2 <\omega^2_0$), the turning point for
SMS waves $r_0$ (for the case $m\neq 0$, the points $r_s $ and $r_0$ appear
only together), the resonance surface for Alfven waves $r_A $, and turning
points $r _ {01} $ and $r _ {02} $ (for $m\neq 0$) appear sequentially in the
system.

 Similarly, for $ \omega _ {SI} ^2 <\omega^2_b $ (where $
\omega^2_b\equiv\bar {\omega} _A^2 (r_b) $) the resonance surface for
SMS waves disappears ($r_s $ is virtually displaced to the right of $r_b
$) from the system. When $ \omega^2_b $ then increases, the
points $r_0$, $r_A $ (for $m=0$ the entire jet becomes a transparency
region) and $r _ {01} $ disappear sequentially. The point $r _ {02}
\rightarrow 0$ when $ \omega^2_b\rightarrow\infty $. This may be imagined
by shifting the function $k_r^2 (r) $ plot, presented in fig.2, farther
to the right through the point $r=r_b $. Thus, depending on the magnitude of
$ \omega/k_z $, both the transparency region and the opacity region for the
waves in question can adjoin the boundary inside the jet.

To picture the behaviour of $k_r^2 (r) $ outside the jet, we will consider
the functions $ \bar {\omega} _ {A1} ^2 (r), \bar {\omega} _ {A2} ^2 (r) $,
presented in fig.3. As follows from the last expression of (\ref {Eq_5}),
when $ \bar {\omega} _ {AII} ^2 <\bar {\omega} _ {SII} ^2$ (where $ \bar
{\omega} _ {AII} ^2\equiv \bar {\omega} _ {A} ^2 (r\rightarrow \infty) $, $
\bar {\omega} _ {SII} ^2\equiv \bar {\omega} _ {S} ^2 (r\rightarrow \infty)
$) \-- the outside jet region adjoining the boundary $r=r_b $ is an opacity
region for the waves in question. When $ \bar {\omega} _ {SII} ^2
<\bar {\omega} _ {AII} ^2 <\bar {\omega} _ {A2b} ^2$ \-- the outside region is
transparent, and for $ \bar {\omega} _ {A2b} ^2 <\bar {\omega} _ {AII} ^2
<\beta ^ * _ {II} $ the transparency region shifts to $r> r_1> r_b $. In
the interval $ \beta ^ * _ {II} <\bar {\omega} _ {AII} ^2 <1 +\beta ^ * _
{II} $ the outside region adjoining the boundary is opaque, but is
transparent when $ \bar {\omega} _ {A1b} ^2 <\bar {\omega} _ {AII} ^2$.

\begin {figure} \label {f3}
{\includegraphics [width=10cm] {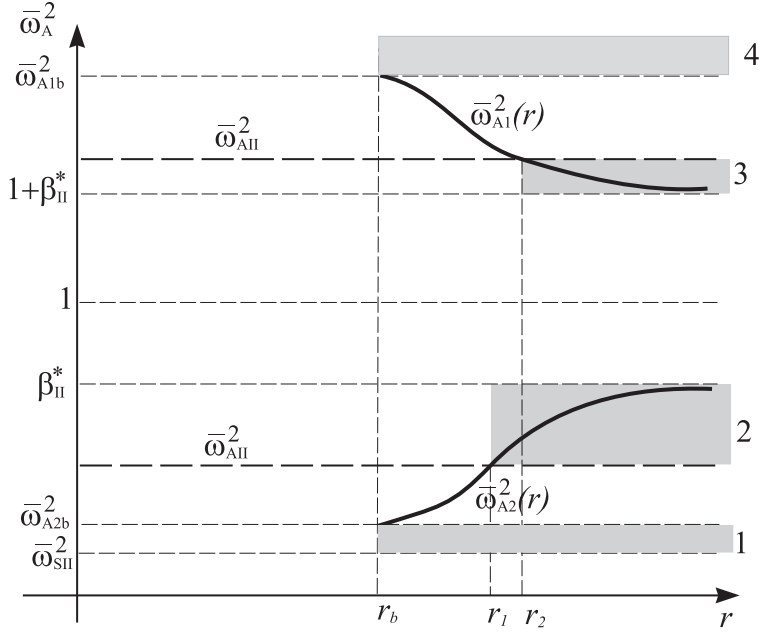}}
  \caption {
Distribution of the functions $ \bar {\omega} _ {A1} ^2 (r), \bar {\omega} _
{A2} ^2 (r) $ and transparency regions (shown in grey) for
SMS (1) and FMS waves (2,3,4), when $r> r_b $. Two possible magnitudes
of $ \bar {\omega} _ {AII} ^2\equiv \bar {\omega} _ {A} ^2
(r\rightarrow \infty) $ (horizontal bold dashed lines) are presented, for
which the turning points $r _ {1,2} $ exist in the regions outside of
the jet.
}
\end {figure}

Thus, depending on the magnitude of $ \bar {\omega} _ {AII} ^2$ determined by
the wave phase velocity given the Doppler displacement
of frequency for the oscillations under study \- $ \bar {\omega}/k_z
$, the boundary can also adjoin both the transparency and the opacity
region outside the jet. As was shown above, the transparency or opacity of
the region adjoining the boundary inside the jet is determined by the magnitude of
the parallel phase velocity $ \omega/k_z $. The solutions describing the
oscillations in the transparency and the opacity regions adjoining the
boundary outside and inside the jet can be combined in all possible
combinations in the matching condition.

It is easy to obtain the matching condition for the solutions on the plasma
jet boundary by integrating the equation (\ref {Eq_4}) in a narrow
interval ($r_b-\varepsilon, r_b +\varepsilon $):

 \begin {equation} \label {Eq_6}
\left.\frac {\rho_0\Omega^2} {k_r^2} \frac {\partial \ln
\zeta} {\partial r} \right | _ {r_b-\varepsilon} = \left.\frac {\rho_0\Omega^2} {k_r^2} \frac {\partial \ln \zeta} {\partial r} \right | _ {r_b +\varepsilon},
\end {equation}
where $ \varepsilon\rightarrow 0$. Using the expressions (\ref {Eq3_b}),
(\ref {Eq3_c}) it is possible to show that (\ref {Eq_6}) is similar to
the requirements for the plasma to be identically displaced
on both sides of the
boundary ($ \zeta _ {r_b-\varepsilon} = \zeta _ {r_b +\varepsilon} $ is the
condition of impermeability) and for the total perturbed pressure to be
sustained
across the boundary ($ (P+B_zB_0/4\pi) _ {r_b-\varepsilon} = (P+B_zB_0/4\pi)
_ {r_b +\varepsilon} $).

Now let us define the boundary conditions for the problem. When
$r\rightarrow 0$ the finite magnitude of the desired solution is
a natural requirement. As to the boundary condition for $r\rightarrow \infty $
its determination is related to the causality principle. In this problem, we
will be interested in solutions to (\ref {Eq_4}) describing unstable modes of
the plasma jet oscillations. For such solutions, oscillations far from
the shift layer are running away from the shear flow that generated them,
according to the causality principle. In other words, the energy flux of
these waves should be directed out of the shift layer.

It should be noted when dealing with unstable oscillations
that the wave vector component $k _ {r} $ in the asymptotics is complex.
For any weak unstable oscillations, it is formally possible to introduce
the concept of waves running from the shift layer, for which $ {\rm
Re} (v _ {gr})> 0$ when $r\rightarrow \infty $, where $v _ {gr} = \partial
\omega/\partial k _ {r} $ is the group velocity with which the
wave energy is transferred over the radius $r $. The
energy conservation law

 \[
\frac {\partial\mathcal {E}} {\partial t} + \frac {1} {r} \frac {\partial} {\partial r} (r\mathbf {v} _ {gr} \mathcal {E}) =0,
\]
 where $ \mathcal {E} $ is wave energy density, quadratic on the oscillation
amplitude, implies that $ {\rm Im} (k _ {r})> 0$ when $r\rightarrow \infty $
for monochromatic unstable oscillations ($ \rm {Im} (\omega)> 0$). This
results in an exponentially decreasing amplitude of oscillations escaping
from the shift layer. A specific expression for the group velocity when
$r\rightarrow \infty $ can be obtained by differentiating the expression (\ref
{Eq_5}) with respect to $ \omega $:

\begin {equation} \label {Eq_7}
v _ {gr} =A _ {II} \frac {1 +\beta ^ * _ {II}} {k_z} {\rm Re} k _ {rII} \frac {\left [\bar {\omega} _ {A} ^2-\bar {\omega} _ {SII} ^2\right] ^2} {\bar {\omega} _ {A} ^3\left [\bar {\omega} _ {A} ^2-2\bar {\omega} _ {SII} ^2\right]}.
\end {equation}
The boundary condition for the wave running from the shift layer when
$r\rightarrow \infty $ has the form

\begin {equation} \label {Eq_9}
\frac {\partial\zeta} {\partial r} =ik _ {rII} \zeta,
\end {equation}
and the sign of $ k _ {rII} = \pm\sqrt {k _ {rII} ^2} \equiv k _ {r}
(r\rightarrow\infty) $ is determined by the requirement $ {\rm Re} (v _
{gr})> 0$.

\section {Structure of MHD oscillations in the jet in the WKB approximation}

In order to understand qualitatively the structure of the oscillations in the
plasma jet, let us consider a problem for the MHD oscillations with
parameters permitting us to use the WKB approximation far from the turning points
and resonance surfaces. We will search for solutions in the neighbourhoods of
these points by linearizing the coefficients in (\ref {Eq_4}) and
subsequently matching the solutions with the solutions obtained in the WKB
approximation. To make a complete picture of the wave field structure, let
us consider oscillations with parameters corresponding to the distribution of
$k_r^2 (r) $ in fig.2, presenting all possible singular and turning points.
Let us consider the structure of the forced and
eigen-oscillations of the plasma jet separately.

\subsection {The structure of forced oscillations in the plasma jet}

Let us consider the case of forced MHD oscillations of the plasma jet,
with a source located at its boundary.

 {\it When $r\rightarrow 0$} equation (\ref {Eq_4}) can be
approximately presented in the form:

 \[
 r^2\zeta''+ r ^\sigma\zeta ' + (k _ {r0} ^2r^2-1) \zeta=0,
\]
where $k _ {r0} ^2\equiv k _ {r} ^2 (r\rightarrow 0) $ (for $m\neq 0$ we have
$k _ {r0} ^2\approx-m^2/r^2$), $ \sigma=1$ when $m=0$ and $ \sigma=3$ when
$m\neq 0$. A solution which is finite when $r\rightarrow 0$ has the form:

 \begin {equation} \label {Eq_10}
 \zeta=C_1\left \{
 \begin {array} {lcr}
   J_1 (\sqrt {k _ {r0} ^2} r),&  for\  m=0, \\
   r ^ {m-1} ,& for\   m\neq 0,
   \end {array} \right.
\end {equation}
where $C_1$ is an arbitrary constant, $J_1 (\sqrt {k _ {r0} ^2} r) $ is the
Bessel function ($J_1 (\sqrt {k _ {r0} ^2} r) \stackrel {r\rightarrow 0}
{\approx} \sqrt {k _ {r0} ^2} r/2$). The subsequent calculations will
concern the case $m\neq 0$.

{\it Let us present the WKB solution in the opacity region $0 <r <r _ {01}
$}, matched with (\ref {Eq_10}), in the form

\begin {equation} \label {Eq_11} \zeta=C_2\frac {(-k_r^2) ^ {1/4}} {\sqrt {\rho_0\Omega^2r}} \exp\left (\int _ {r _ {01}} ^r\sqrt {-k_r^2} dr '\right),
\end {equation}
where $C_2$ is an arbitrary constant. Matching (\ref {Eq_10}) to (\ref
{Eq_11}) we obtain a relation for the constants $C_1=C_2\exp
{(\psi_1)}/\bar {r} ^m\sqrt {\rho_0\Omega^2_0/m} $, where $ \psi_1 =\int _ {r
_ {01}} ^ {\bar {r}} \sqrt {-k_r^2} dr $, and $ \bar {r}> 0$ is an arbitrary
point in the neighbourhood of $r=0$.

{\it Solution in the neighbourhood of the turning point $r=r _ {01} $.}
Let us differentiate (\ref {Eq_4}) with respect to $r $, introduce the notation
$u = (1/k_r^2) \partial\zeta/\partial r $ and, linearizing $k_r^2\approx
\xi_1/a _ {1} ^2$ close to $r=r _ {01} $ (where $a_1 ^ {-3} = (\partial
k_r^2/\partial r) _ {r=r _ {01}} $, $ \xi_1 = (r-r _ {01})/a_1$), will yield
an equation

\begin {equation} \label {Eq_12}
    \frac {\partial^2u} {\partial \xi_1^2} + \xi_1u=0.
\end {equation}
Its solution matched to (\ref {Eq_11}) has the form
 \begin {equation} \label {Eq_12_1} u=C_3Ai (-\xi_1),
\end {equation}
where $Ai (z) $ is the Airy function. Matching this solution to
(\ref {Eq_11}) yields a relation for the constants $C_3=2C_2\sqrt {\pi a_1/\rho
_ {01} \Omega^2_1r _ {01}} $. Hereafter the subscripts $ _0, _1, _2, _A, _S $
denote the parameters at the corresponding points $r_0, r _ {01}, r _ {02},
r_A, r_S $.

{\it The WKB solution in the transparency region $r _ {01} <r <r _ {02} $},
matched to (\ref {Eq_12_1}) has the form

\begin {equation} \label {Eq_13} u =\frac {C_4} {\sqrt {\rho_0\Omega^2k_rr}} \sin\left (\int _ {r _ {01}} ^rk_rdr ' + \frac {\pi} {4} \right),
\end {equation}
where $C_4=2C_2$.

{\it The solution in the neighbourhood of the turning point $r=r _ {02} $} may
be obtained using linearization $k_r^2\approx-\-\xi_2/a _ {2} ^2$
($a_2 ^ {-3} = (\partial k_r^2/\partial r) _ {r=r _ {02}} $, $ \xi_2 = (r-r _
{02})/a_2$) and obtaining an equation similar to (\ref {Eq_12}) accurate
within the substitution $ \xi_1\rightarrow-\xi_2$. Its solution matched to
(\ref {Eq_13}) is

\begin {equation} \label {Eq_14}
u=C_5Ai (\xi_2) +C_6Bi (\xi_2),
\end {equation}
where $Ai (z), Bi (z) $ are the Airy functions. We have the constants
$C_5=C_4\sqrt {\pi a_2/\rho _ {02} \Omega^2_2r _ {02}} \sin {\psi_2} $,
$C_6=C_4\sqrt {\pi a_2/\rho _ {02} \Omega^2_2r _ {02}} \cos {\psi_2} $, where
$ \psi_2 =\int _ {r _ {01}} ^ {r _ {02}} k_rdr $. If $ \psi_2 =\pi (n+1/2) $
($n=0,1,2... $) and $C_6=0$, then to the right of $r _ {02} $ in the
opacity region we have a solution exponentially decreasing in amplitude (the
asymptotics of the functions when $z\rightarrow\infty $: $Ai (z) = \exp [-
(2/3) z ^ {3/2}]/2\sqrt {\pi\sqrt {z}}, \ \ Bi (z) = \exp [(2/3) z ^
{3/2}])/\sqrt {\pi\sqrt {z}} $). This condition means that the eigen-waveguide
mode of FMS waves is captured in the transparency region ($r _ {01}, r _ {02}
$). As to the case $m=0$, a solution similar to (\ref {Eq_13}) is obtained
for the transparency region ($0, r_A $).

{\it The WKB solution in the opacity region $r _ {02} <r <r _ {A} $} will
be presented in the form:

\begin {equation} \label {Eq_15}
    \zeta =\frac {(-k_r^2) ^ {1/4}} {\sqrt {\rho_0\Omega^2r}} \left [C_7\exp\left (\int _ {r _ {A}} ^r\sqrt {-k_r^2} dr '\right) +C_8\exp\left (-\int _ {r _ {A}} ^r\sqrt {-k_r^2} dr '\right) \right].
\end {equation}
If an eigen mode propagates in the waveguide ($r _ {01}, r _ {02} $), there
is only a solution exponentially decreasing in amplitude (i.e. $C_7=0$) in
the opacity region $r _ {02} <r <r _ {A} $. That case will be dealt with in
the following Section. For the non-eigenmodes whose source are
the plasma jet boundary oscillations, an exponentially growing solution
exists in the opacity region, against the background of which, to stick to an
exponentially decreasing solution means to exceed the accuracy. In that case
matching the solutions (\ref {Eq_14}) and (\ref {Eq_15}) yields
$C_7=C_6\exp (\psi_3) \sqrt {\rho _ {02} \Omega^2_2r _ {02}/\pi a _ {2}} $,
where $ \psi_3 =\int _ {r _ {02}} ^ {r_A} \sqrt {-k_r^2} dr $.

{\it Solution near the resonance surface $r=r_A $.} When $r\rightarrow
r_A $ we have $k_r^2\approx-m^2/r_A^2$. Let us linearize $
\Omega^2\approx k_z^2A^2 (r_A) [(r-r_A)/a_A-2i\gamma/k_zA (r_A)] $ close to
$r=r_A $, where $a_A = (\partial \ln (A ^ {2})/\partial r) ^ {-1} _
{r=r_A} $ is the characteristic scale of $A (r) $ variation. Here the
imaginary part of the frequency $ \gamma\equiv {\rm Im} (\omega) $, necessary
for regularizing the solution near the singular point, is represented in an
explicit form. From (\ref {Eq_4}), we have the equation

\begin {equation} \label {Eq_16}
    \frac {\partial} {\partial \xi_A} (\xi_A-i\varepsilon_A) \frac {\partial \zeta} {\partial \xi_A} - (\xi_A-i\varepsilon_A) \zeta=0,
\end {equation}
where $ \xi_A=m (r-r_A)/r_A $, $ \varepsilon_A=ma_A\gamma/k_zr_AA (r_A) $ (we
assume $ \varepsilon_A\ll 1$). The solution of (\ref {Eq_16}) matched to
(\ref {Eq_15}) is
\begin {equation} \label {Eq_17}
    \zeta=C_9K_0 (-\-\xi_A+i\varepsilon_A),
\end {equation}
where $K_0 (z) $ is the modified Bessel function, $C_9=C_7m\sqrt
{2a_A/\pi\rho _ {0A} r_A} / (k_zA (r_A) r_A) $. When $r\rightarrow r_A $ the
solution has a well known logarithmic singularity

\[
 \zeta\approx-C_9\ln {(-\-\xi_A+i\varepsilon_A)},
\]
which corresponds to the resonant Alfven wave.

 {\it WKB solution in the opacity region $r _ {A} <r <r_0$} will
be presented in the form

\begin {equation} \label {Eq_18}
    \zeta=C _ {10} \frac {(-k_r^2) ^ {1/4}} {\sqrt {\rho_0\Omega^2r}} \exp\left (\int _ {r _ {A}} ^r\sqrt {-k_r^2} dr '\right).
\end {equation}
Matching it to the solution (\ref {Eq_17}) relates the constants : $C _
{10} =C_7$.

{\it Solution in the neighbourhood of the turning point $r=r _ {0} $.}
Let us differentiate (\ref {Eq_4}) with respect to $r $, introduce the notation
$u = (1/k_r^2) \partial\zeta/\partial r $, and, linearizing $k_r^2\approx
\xi_0/a_0^2$ close to $r=r _ {0} $ (where $a_0 ^ {-3} = (\partial
k_r^2/\partial r) _ {r=r _ {0}} $, $ \xi_0 = (r-r _ {0})/a_0$) yields an
equation similar to (\ref {Eq_12}), accurate within the substitute $
\xi_1\rightarrow\xi_0$. Its solution matched to (\ref {Eq_18}) has the form
 \begin {equation} \label {Eq_19}
    u=C _ {11} Ai (-\xi_0),
\end {equation}
 where $C _ {11} =2C _ {10} \exp (\psi_4) \sqrt {\pi a_0/\rho _ {00}
\Omega^2_0r_0} $, $ \psi_4 =\int _ {r_A} ^ {r _ {0}} \sqrt {-k_r^2} dr $.

{\it WKB solution in the transparency region $r_0 <r <r _ {S} $} matched
to (\ref {Eq_19}) may be represented in the form

\begin {equation} \label {Eq_20} \zeta=C _ {12} \sqrt {\frac {k_r} {\rho_0\Omega^2r}} \cos\left (\int _ {r _ {0}} ^rk_rdr ' + \frac {\pi} {4} \right),
\end {equation}
where $C _ {12} =-C _ {11} \sqrt {\rho _ {00} \Omega^2_0r _ {0}/\pi a _ {0}}
$.

{\it Solution near the resonance surface $r=r_S $.} Let us linearize the
coefficient of the higher derivative in (\ref {Eq_4}) representing $k_r ^
{-2} \approx a_s^2\xi_s $, where $ \xi_s = (r-r_S)/a_s $, $a_s = (-\partial
k_r ^ {-2}/\partial r) _ {r=r_s} $ is the characteristic scale of variation
of $k_r ^ {-2} $, close to $r=r_S $. Then, close to
$r=r_S $, equation (\ref {Eq_4}) can be represented in the form

\begin {equation} \label {Eq_21} \frac {\partial} {\partial \xi_s} (\xi_s+i\varepsilon_s) \frac {\partial \zeta} {\partial \xi_s}-\zeta=0,
\end {equation}
where $ \varepsilon_s=ma_s\gamma/k_zr_SC_s (r_S) $ is the regularized factor
related to the imaginary part of the frequency. Its solution matched to
(\ref {Eq_20}) is

\begin {equation} \label {Eq_22}
    \zeta=C _ {13} I_0 (2\sqrt {\xi_s+i\varepsilon_s}) +C _ {14} K_0 (2\sqrt {\xi_s+i\varepsilon_s}),
\end {equation}
$I_0 (z), K_0 (z) $ are the modified Bessel functions. Using their asymptotic
representations when $ \xi_s\rightarrow-\infty $ we find a relation between
the integration constants: $C _ {13} =-iC _ {12} \sqrt {\pi a_s/\rho _ {0s}
\Omega^2_sr _ {s}} \exp {(i\psi_5)} $, $C _ {14} =2C _ {12} \sqrt {\pi
a_s/\rho _ {0s} \Omega^2_sr _ {s}} \cos ({\psi_5}) $, where $ \psi_5 =\int _
{r _ {0}} ^ {r _ {S}} k_rdr $. If $ \psi_5 =\pi (n+1/2) $ ($n=0,1,2... $), we
have $C _ {14} =0$ and the magnetosonic resonance disappears. In all the other
cases, when $r\rightarrow r_S $, there exists a solution with a logarithmic
singularity

\[
 \zeta =-\frac {C _ {14}} {2} \ln {(\xi_s+i\varepsilon_s)},
\]
which corresponds to the resonant SMS wave.

{\it WKB solution in the opacity region $r _ {S} <r <r_b $} for the
oscillations whose amplitude decreases from the boundary into the plasma jet,
will be represented in the form

\begin {equation} \label {Eq_23}
    \zeta=C _ {15} \exp (\psi_6) \frac {(-k_r^2) ^ {1/4}} {\sqrt {\rho_0\Omega^2r}} \exp\left (\int _ {r _ {b}} ^r\sqrt {-k_r^2} dr '\right),
\end {equation}
where $ \psi_6 =\int _ {r _ {S}} ^ {r_b} \sqrt {-k_r^2} dr $. Matching it
to the solution (\ref {Eq_22}) relates the constants: $C _ {15} =C _ {13}
\sqrt {a_s/\pi\rho _ {0s} \Omega^2_sr_s}/2$.

\subsection {Structure of the eigen-modes of MHD oscillations in the plasma
jet}

The wave field of an eigen-mode propagating in the FMS waveguide ($r _ {01}, r
_ {02} $) has the same structure in the opacity region $0 <r\le r _ {02} $ as
in the previous case, described by expressions (\ref {Eq_10}),
(\ref {Eq_11}) and (\ref {Eq_12_1}); in the transparency regions $r _
{01} <r <r _ {02} $  by  expression (\ref {Eq_13}). Near the
turning point $r _ {02} $, however, we have $C_6=0$ in the solution (\ref
{Eq_14}), while in the opacity region $r _ {02} <r <r _ {A} $ this solution has
the amplitude decreasing from the turning point into the opacity region. The
WKB solution in the opacity region $r _ {02} <r <r _ {A} $ looks like (\ref
{Eq_15}) where $C_7=0$. Matching this solution to (\ref {Eq_14})
yields $C_8=C_5\exp (-\psi_3) (-1) ^ {n+1}/2$, where $n=1,2,3... $ is the number
of the eigen harmonic propagating in the FMS waveguide. Near the resonance
surface $r=r_A $ this solution is matched to the solution

\begin {equation} \label {Eq_24_1} \zeta=C_9K_0 (\xi_A-i\varepsilon_A),
\end {equation}
where $C_9=iC_8m\sqrt {2a_A/\pi\rho _ {0A} r_A} / (k_zA (r_A) r_A) $,
having a logarithmic singularity on the resonance surface. In the opacity
region $r _ {A} <r <r_0$, the WKB solution will be presented in the form

\begin {equation} \label {Eq_25_1}
    \zeta=C _ {10} \frac {(-k_r^2) ^ {1/4}} {\sqrt {\rho_0\Omega^2r}} \exp\left (-\int _ {r _ {A}} ^r\sqrt {-k_r^2} dr '\right).
\end {equation}
Matching it to (\ref {Eq_24_1}) yields $C _ {10} =iC_8$.

Near the turning point $r=r_0$, the full solution for the
function $u = (1/k_r^2) \partial\zeta/\partial r $ has the form

\begin {equation} \label {Eq_26_1} u=C _ {11} Ai (-\xi_0) +C _ {12} Bi (-\xi_0).
\end {equation}
To correctly match this solution to the WKB solution
left and right of the turning point, it is necessary to specify its
behaviour in the opacity region $r> r_S $, to the right of the resonance
surface $r=r_S $. Since we are dealing with an eigen-mode, we will require
the amplitude of this solution to decrease into the opacity region $r> r_S $.
The solution in the transparency region $r_0 <r <r_S $ has the form of a
wave coming to the resonance surface

\begin {equation} \label {Eq_27_1}
    \zeta=C _ {13} \sqrt {\frac {k_r} {\rho_0\Omega^2r}} \exp\left (i\int _ {r _ {0}} ^rk_rdr '\right).
\end {equation}
Near the resonance surface $r=r_S $ it is matched to the solution

\begin {equation} \label {Eq_28_1} \zeta=C _ {14} K_0 (2\sqrt {\xi_s+i\varepsilon_s}),
\end {equation}
which last continues into the opacity region $r> r_S $ by a WKB solution of
the form

\begin {equation} \label {Eq_29_1}
    \zeta=C _ {15} \frac {(-k_r^2) ^ {1/4}} {\sqrt {\rho_0\Omega^2r}} \exp\left (-\int _ {r _ {S}} ^r\sqrt {-k_r^2} dr '\right).
\end {equation}

The constants in the solutions (\ref {Eq_25_1}), (\ref {Eq_26_1}), (\ref
{Eq_27_1}), (\ref {Eq_28_1}) and (\ref {Eq_29_1}) are related:
$C _ {12} =-iC _ {11} =-C _ {10} \exp ({-\psi_4}) \sqrt {\pi a _
{0}/\rho _ {00} \Omega^2_0r _ {0}} $, $C _ {13} =-C _ {10} \exp
({-\-\psi_4+i\pi/4}) $, $C _ {14} =2C _ {13} \exp (i\psi_5-i\pi/4)/\sqrt {\pi
a_s\rho _ {0s} \Omega^2_sr _ {s}} $, $C _ {15} =C _ {13} \exp
(i\psi_5-i\pi/4) $. The eigen-mode propagates in the FMS waveguide and is
simultaneously absorbed on the resonance surfaces for the Alfven and
SMS waves. The entire wave energy reaching the resonance surface for
SMS waves is completely absorbed in the neighbourhood of the surface
\cite{LeonKoz 2009}. Thus, the Q factor of oscillations in such a waveguide is less than
unity even in the absence of waves escaping from the waveguide.

\section {WKB calculation of the increment of shear flow instability on the
plasma jet boundary}

Let us match the internal solution for forced oscillations, obtained by the
WKB approximation in the previous Section, to the external solution
describing the structure of oscillations outside the jet. We will consider
the plasma jet boundary as a tangential discontinuity when $r=r_b $. Note
that in this approximation, describable as local, the dispersion properties
of the oscillations are determined by the parameters of the
immediately adjoining medium inside and outside of the jet boundary. In this
case the result does not depend on the variation of medium properties far
from the tangential discontinuity. The matching condition (\ref {Eq_6})
lets us obtain the dispersion equation in the form

\begin {equation} \label {Eq_24}
 b\frac {c^2-1} {c^2-M^2_A} = \left \{
 \begin {array} {lcl}
   -\sqrt {k^2 _ {rI}/k^2 _ {rII}},& for & {\rm Re} (k^2 _ {rI}), {\rm Re} (k^2 _ {rII}) <0, \\i\sqrt {-k^2 _ {rI}/k^2 _ {rII}},& for & {\rm Re} (k^2 _ {rI}) <0, {\rm Re} (k^2 _ {rII})> 0, \\ -\-\cot\psi\sqrt {-k^2 _ {rI}/k^2 _ {rII}}, & for & {\rm
Re} (k^2 _ {rI})> 0, {\rm Re} (k^2 _ {rII}) <0, \\ i\cot\psi\sqrt {k^2 _ {rI}/k^2 _ {rII}}, & for & {\rm Re} (k^2 _ {rI}), {\rm Re} (k^2 _ {rII})> 0, \\ \end {array}
 \right.
\end {equation}
  where the subscripts $ _I, _ {II} $ denote the values at point $r=r_b $
on the inner and outer side of the jet, respectively, $b=B^2 _ {0I}/B^2 _
{0II} $, $c =\bar {\omega}/k_zA_I $, $M_A=v _ {0II}/A_I $ is the Mach number
defined by the Alfven speed $A_I $, $ \psi =\int _ {\bar {r} _ {0}} ^ {r _
{b}} k_rdr +\pi/4$ is phase incursion in the transparency region ($ \bar {r}
_ {0}, r _ {b} $) adjoining the jet boundary from the inside ($ \bar {r} _
{0} =r _ {0} $ for the SMS waves, $ \bar {r} _ {0} =r _ {01} $ for
the FMS waves when $m\neq 0$, $ \bar {r} _ {0} =0$ for the FMS waves when
$m=0$). In the same notations

\begin {eqnarray*}
k^2 _ {rI} &=& k_z^2\left (\frac {c^4} {c^2 (1 +\beta ^ * _ I)-\beta ^ * _ I}-1-\kappa_m^2\right), \\
k^2 _ {rII} &=& k_z^2\left (\epsilon ^ {-2} \frac {(c-M_A) ^4} {(c-M_A) ^2 (1 +\beta ^ * _ {II})-\-\epsilon^2\beta ^ * _ {II}}-1-\kappa_m^2 \right),
\end {eqnarray*}
where $ \beta ^ * _ {I, II} =S _ {I, II} ^2/A _ {I, II} ^2$, $
\kappa_m=m/k_zr_b $, $ \epsilon=A _ {II}/A_I $ (we assume $A _ {II} \ll A_I
$). We will employ the perturbation technique using small-parameter ($
\epsilon\ll 1$) expansion to search for the solution of the dispersion
equation (\ref {Eq_24}), assuming

\begin {equation} \label {Eq_24_1} c=c_0 +\epsilon c_1 +...
\end {equation}
In the zeroth order of the perturbation theory, we have $c_0=M_A $. In the
first order of the perturbation theory, squaring the left and right parts of
(\ref {Eq_24}) produces the equation for $c_1$:

\begin {equation} \label {Eq_25}
    \bar {b} ^2 (M_A^2-1) ^2\left (
    \frac {c_1^4} {c_1^2 (1 +\beta ^ * _ {II})-\beta ^ * _ {II}}-1-\kappa_m^2\right) = \pm (c_1^2-1) ^2k^2 _ {rI0}
\end {equation}
where $k^2 _ {rI0} \equiv k^2 _ {rI} (c_0=M_A) $. The plus sign on the right
hand and $ \bar {b} =b $ corresponds to $ {\rm Re} (k^2 _ {rI}) <0$, the
minus sign and $ \bar {b} =b\tan (\psi +\pi/4) $ corresponds to $ {\rm Re}
(k^2 _ {rI})> 0$. Equation (\ref {Eq_25}) is of the sixth order with respect
to $c_1$ and its solution can be sought for numerically. However when $ |c_1
|\gg 1$ (but $ \epsilon|c_1 |\ll c_0$) it can be approximately reduced to the
biquadratic equation

\begin {equation} \label {Eq_26}
    c_1^4\mp
    c_1^2\frac {\bar {b} ^2 (M_A^2-1) ^2} {k^2 _ {rI0} (1 +\beta ^ * _ {II})} \pm\frac {\bar {b} ^2} {k^2 _ {rI0}} (1 +\kappa_m^2) (M_A^2-1) ^2=0.
\end {equation}
The solution of (\ref {Eq_26}) for $ {\rm Re} (k^2 _ {rI}) <0$ has the form

\begin {equation} \label {Eq_27} c_1^2 =\frac {b^2 (M_A^2-1) ^2} {2k^2 _ {rI0} (1 +\beta ^ * _ {II})} \pm\sqrt {\frac {b^4 (M_A^2-1) ^4} {4k^4 _ {rI0} (1 +\beta ^ * _ {II}) ^2} \frac {b^2 (M_A^2-1) ^2 (1 +\kappa_m^2)} {k^2 _ {rI0}}}.
\end {equation}

Obviously, the condition $ |c_1 |\gg 1$ is satisfied when $b\gg 1$ and $
|M^2_A-1 |\stackrel {>} {\sim} 1$. The value

\[
k^2 _ {rI0} =k_z^2\left (
\frac {M_A^4} {M_A^2 (1 +\beta ^ * _ {I})-\beta ^ * _ {I}}-1-\kappa_m^2\right) \]
is real and, hence, when $k^2 _ {rI0}> 0$ ($c_1^2> 0$) there are no
unstable oscillations, but when $k^2 _ {rI0} <0$, we obtain
the solution for an unstable
mode if we choose to have the minus sign before the radical in (\ref
{Eq_27}). It is easy to check that $k^2 _ {rI0} <0$ when $M_A <M_0$ and $M_1
<M_A <M_2$, where $M^2_0 =\beta ^ * _ {I} / (1 +\beta ^ * _ {I}) $, and $M^2
_ {1,2} $ are the roots of the biquadratic (with respect to $M_A $) equation
$k^2 _ {rI0} =0$:

\[
 M^2 _ {1,2} = \frac {(1 +\kappa_m^2) (1 +\beta ^ * _ {I})} {2} \pm \sqrt {\frac {(1 +\kappa_m^2) ^2 (1 +\beta ^ * _ {I}) ^2} {4}-\beta ^ * _ {I} (1 +\kappa_m^2)}.
  \]
When $ \beta ^ * _ {I} \ll 1$ we have $M^2 _ {1} \approx M^2 _ {0} +M^4 _
{0}/M^2_2 <1$ and $M^2 _ {2} \approx (1 +\kappa_m^2) (1 +\beta ^ * _ {I})>
1$.

As follows from the second equation of (\ref {Eq_24}), the value $c_1^2$
cannot be real when $ {\rm Re} (k^2 _ {rII})> 0$ (which corresponds to a
transparency region outside of the jet), which contradicts the solution (\ref
{Eq_27}) when $k^2 _ {rI0} <0$. In this case there are no unstable
oscillations. In the case $ {\rm Re} (k^2 _ {rII}) <0$, corresponding to
the first equation of (\ref {Eq_24}), the signs in the left- and right-hand
parts of equation (\ref{Eq_24}) are only in agreement when $M_A> 1$.
Hence, when $ {\rm Re} (k^2 _ {rI}) <0$ the unstable oscillations are driven
on the plasma jet boundary for the flux parameter range

\begin {equation} \label {Eq_28}
    1 <M_A <M_2.
\end {equation}

The solution of (\ref {Eq_26}) when $ {\rm Re} (k^2 _ {rI})> 0$ far from the
poles ($ \bar {b} ^2 =\infty $) and zeros ($\bar {b} ^2=0$) of the function
$\bar {b} =b\tan (\psi +\pi/4) $ has the form

\begin {equation} \label {Eq_29}
c_1^2 =-\frac {\bar {b} ^2 (M_A^2-1) ^2} {2k^2 _ {rI0} (1 +\beta ^ * _ {II})} \pm\sqrt {\frac {\bar {b} ^4 (M_A^2-1) ^4} {4k^4 _ {rI0} (1 +\beta ^ * _ {II}) ^2} + \frac {\bar {b} ^2 (M_A^2-1) ^2 (1 +\kappa_m^2)} {k^2 _ {rI0}}}.
\end {equation}
Unstable solutions are obtained when $k^2 _ {rI0}> 0$, which corresponds to
the parameter ranges $M_0 <M_A <M_1$ and $M_A> M_2$. The same as in the case
$k^2 _ {rI0} <0$, the solutions corresponding to a transparency region
outside of the jet ($ {\rm Re} (k^2 _ {rII})> 0$) describe steady-state
oscillations only. Analysis of the signs in the left- and right-hand part of
the third equation (\ref {Eq_24}) for the opaque external region ($ {\rm Re}
(k^2 _ {rII}) <0$) shows that it is only the positive (right of the solutions
of equation $ \tan (\psi (c _ {0n}) + \pi/4) =0$, where $n=1,2,3... $)
branches of the functions $ \bar {b} (M_A) =b\tan (\psi (M_A) + \pi/4)> 0$
that correspond to the unstable solutions when $M_A> M_2> 1$, with only the
negative (left-hand) branches: $ \bar {b} (M_A) =b\tan (\psi (M_A) + \pi/4)
<0$ corresponding to them when $M_0 <M_A <M_1 <1$. The first of these
parameter ranges ($M_A> M_2$) corresponds to an FMS wave transparency region
ajoining the plasma jet boundary from the inside, while the second range
($M_0 <M_A <M_1$) to an SMS wave transparency region. It is possible to show
that the solutions describe only stable oscillations (Im $ (c) <0$) when
approaching the poles and the zeros of the function $ \bar {b} (M_A) $.

Fig.~4 is an example of a numerical solution of the dispersion equation
(\ref {Eq_24}) for the azimuthal harmonic $m=1$, parallel component of the
wave vector $k_zr_b=2$ and the following parameters of the medium: $
\epsilon=A _ {II}/A_I=0.08$, $ \beta_I ^ * = 0.05$, $b=B _ {0I} ^2/B _ {0II}
^2=16$. Such a parameter set is characteristic of the Earth's magnetotail,
which the solar wind flows around. Notably, no eigen-mode does exist in the
waveguide for the SMS waves. Therefore the plasma jet is stable when $M_A
<1$. When $M_A> 1$, the ranges of the Mach numbers $M_A $ for unstable modes
correspond to the interval (\ref {Eq_28}) and to the above solutions of
equation (\ref {Eq_29}), corresponding to the positive branches
of the functions: $ \bar {b} (M_A) =b\tan (\psi (M_A) + \pi/4)> 0$. Each of
these roots corresponds to one of the eigen harmonics of the FMS waveguide
adjoining the plasma jet boundary from the inside. When $M_A $ increases,
higher and still higher harmonics become unstable. Fig.5 displays no
appreciable $M_A $ dependence of the maximum magnitudes of the oscillations
increment, but the ranges of the unstable oscillations decreases noticeably
when $M_A $ (and eigen harmonic number $n $) increases.
\begin {figure} \label {f4}
{\includegraphics [width=10cm] {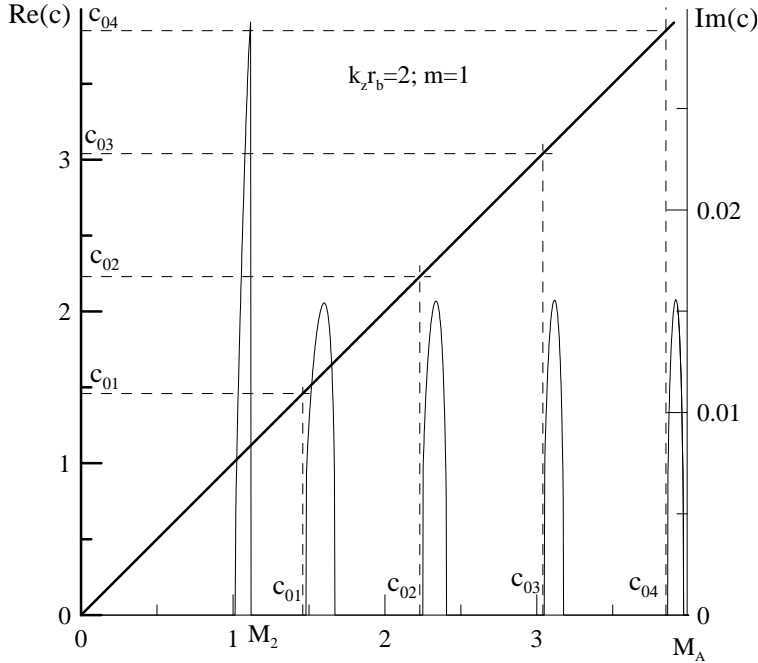}} \caption {Mach number $M_A
$ dependence of the frequencies (Re ($c $), bold line) and of the
increment (Im ($c $), thin lines) of the plasma jet eigen-oscillations
driven on its boundary having the form of a tangential
discontinuity. The solution is obtained in the WKB approximation on the
radial coordinate $r $ for the following parameters of the medium: $
\epsilon=0.08$, $ \beta_I ^ * = 0.05$, $b=16$. $c _ {01,02,03,04} $ are the
roots of the dispersion equation $ \tan (\psi (c _ {0n}) + \pi/4) =0$
defining the eigen-frequency $c _ {0n} $ of the waveguide adjoining
the plasma jet boundary from the inside.} 
\end {figure}

\section {Instability of a plasma jet with a smooth boundary}

Let us now consider a problem of the stability of inhomogeneous plasma jet
with a boundary in the form of a smoothly varying transition layer. We will
not suggest the applicability of the WKB method searching for a solution,
which will allow us to use the following results to oscillations with any
spatial structure. In this case a solution to (\ref {Eq_4}) may only be
found numerically. For a convenient search of a numerical solution and
comparison with the above results, we will rewrite (\ref {Eq_4}) in the
dimensionless form

\begin {equation} \label {Eq_30} \frac {\partial} {\partial x} \, \, \frac {\tilde {b} ^2 (x) (\bar {\omega} _A^2 (x)-1)} {x\kappa^2 (x)} \, \, \frac {\partial x\zeta} {\partial x} + (k_zr_b) ^2\tilde {b} ^2 (x) (\bar {\omega} _A^2 (x)-1) \zeta=0,
\end {equation}
where $x=r/r_b $, $ \bar {\omega} _A (x) = [c-M_A\tilde {v} _0
(x)]/\tilde {v} _A (x) $, $ \tilde {v} _A (x) =A (x)/A_I $, $ \tilde {v} _0
(x) =v_0 (x)/v _ {0I} $, $ \tilde {b} (x) =B^2 _ {0} (x)/B^2 _ {0I} $, \[
\kappa^2 (x) = \frac {\bar {\omega} _A^4} {\bar {\omega} _A^2 (x) (1 +\beta ^
* (x))-\beta ^ * (x)}-1-\frac {\kappa_m^2} {x^2}, \] $ \beta ^ * (x) =A^2
(x)/S^2 (x) $. The profiles of the shear flow velocity $ \tilde {v} _0 (x) $,
Alfven speed $ \tilde {v} _A (x) $ and of the square of the magnetic field
strength $ \tilde {b} (x) $ will be simulated by the following functions:

\begin {eqnarray*}
    \tilde {v} _0 (x) = \frac {1} {2} \left [1 +\tanh\frac {x-1} {\Delta} \right], \\
 \tilde {v} _A (x) = \frac {1} {2} \left [\epsilon +\epsilon_0 + (1-\epsilon_0) \sqrt {x} + (\epsilon + \epsilon_0 - (1-\epsilon_0) \sqrt {x}) \tanh\frac {x-1} {\Delta} \right], \\
 \tilde {b} (x) = \frac {1} {2} \left [1+b ^ {-1} - (1-b ^ {-1}) \tanh\frac {x-1} {\Delta} \right],
\end {eqnarray*}
where $ \Delta =\Delta_r/r_b $, $ \epsilon=A _ {II}/A _ {I} $, $ \epsilon_0=A
(0)/A _ {I} $, $b=B^2 _ {0I}/B^2 _ {0II} $, and the function $ \beta ^ * (x)
$ will be determined from the equilibrium condition of the plasma
configuration (\ref {Eq_2}):

\[
  \beta ^ * (x) = \frac {\beta ^ * _ I} {\tilde {b} (x)} + \frac {\gamma} {2} \left (\frac {1} {\tilde {b} (x)}-1\right).
  \]

The numerical calculations involved the following magnitudes of
the medium parameters: $ \Delta=0.066$, $b=16$, $ \epsilon_0=0.016$, $
\epsilon=0.008$, $ \beta ^ * _ I=0.05$. What we were solving was the
boundary-value problem of searching the phase velocity of oscillations
(the $c $ parameter, in dimensionless variables) satisfying the
boundary conditions (\ref {Eq_9}) when $x\rightarrow\infty $ and (\ref
{Eq_10}) $x\rightarrow 0$. Fig. 5 displays the results of numerical
calculations of the increment of unstable oscillations for the azimuthal
harmonic $m=1$ and parallel wave number $k_zr_b=2$.
Comparison with fig.4, presenting the solution of the same problem in the
local approximation for oscillations of a sharp plasma-jet boundary,
demonstrates essential different distributions of the oscillation
increment.

\begin {figure} \label {f5}
{\includegraphics [width=10cm] {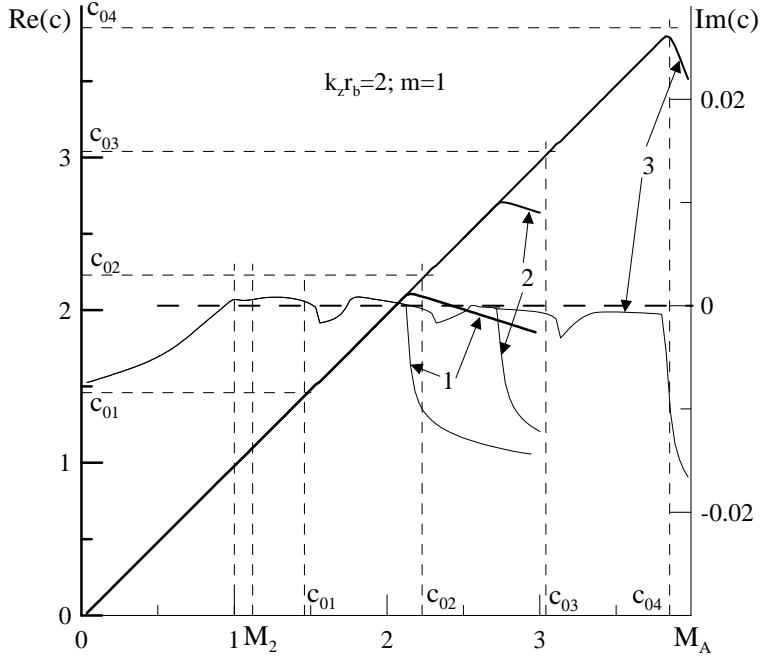}}
\caption {Mach number $M_A $ dependence of the frequencies (Re ($c $),
bold line) and of the increment (Im ($c $), thin lines) of the
eigen-oscillations of the plasma jet with a boundary in the form of a smoothly
varying transition layer with characteristic thickness $
\Delta\equiv\Delta_r/r_b=0.066$. This is a numerical solution of (\ref
{Eq_30}) for the same parameters of the medium ($ \epsilon=0.08$, $ \beta_I ^ * =
0.05$, $b=16$) as in Fig.4.
}
\end {figure}

First, it should be noted that the solution for the plasma jet with a smooth
boundary in the $c (M_A) $ plot represents a "bundle" of curves when $M_A>
M_2$. These curves diverge from the basic value of $c\approx M_A $ (the
zero-approximation solution in the previous Section) when they pass through
the eigenvalues $Re (c) =c _ {0n} $ corresponding to the values of
the eigen-mode frequency of the FMS waveguide adjoining the jet boundary.
This poses considerable difficulties for a numerical search of the
required solution. The solutions were found by numerically
integrating equation (\ref {Eq_30}) (using the Runge-Kutt method) and by
searching the $c $ values (using the Newton method) corresponding to the
boundary conditions (\ref {Eq_9}), (\ref {Eq_10}). The following has proved
to be an optimal technique for calculating the branch corresponding to the $n
$\--th eigen-mode. The calculation starts from the maximum magnitude $M _
{A, max} $ towards $M_A=0$ in the calculating grid. The value of $M _ {A,
max} $ should be chosen as somewhat higher than $c _ {0n} $, which will allow
us to determine numerically the root with the maximum gradient of $ {\rm
Im} (c) $ corresponding to the $n $\--th eigen-mode. Note that it is only
solutions with $ {\rm Im} (c)> 0$ (for the boundary conditions (\ref
{Eq_9}), they correspond to waves running from the plasma jet boundary) that
have a physical sense. Solutions with $ {\rm Im} (c) <0$ should only be
regarded as analytic continuations of solutions with $ {\rm Im}
(c)> 0$. Integration is in increments allowing us to remain on the
already calculated branch until the previous value $c _ {0n-1} $. After the
point of crossing the imaginary parts $ {\rm Im} (c _ {0n}) = {\rm Im} (c _
{0n-1}) $, we switch to the previous branch with a solution corresponding to
the $(n-1)$\--th harmonic.

Fig.5 presents the solutions in the range $0 <M_A <4$ including the
solutions for harmonics $n=1,2,3$. Comparison with fig.4 shows
a manifold decrease of the increment of unstable oscillations. Only
the few first harmonics ($n=1,2$ in our case) remain unstable. This is
explained by the smoothing of the boundary layer and by competition between
the dissipation mechanism of oscillations on resonance surfaces and the
mechanism of shear flow instability. Aditionally, the points of the
eigen-mode phase velocity $c _ {0n} $ are displaced (fig.5 shows the same
points $ c _ {0n} $ and $M_2$ as those in fig.4, obtained in the WKB
approximation), the first region of unstable oscillations extends.

Fig.6a demonstrates the spatial structure of unstable oscillations close to
the second harmonic $n=2$. As was to be expected from the analysis of the WKB
solution, the region outside of the jet is opaque for the unstable
oscillation mode. Resonance surfaces for the Alfven and SMS waves determined,
respectively, by the equations $ \bar {\omega} _A (r_A) = \pm 1$ and $ \bar
{\omega} _A (r_S) = \pm \bar {\omega} _s (r_S) $ are located in the region of
the transition layer at the plasma jet boundary.
\begin {figure} \label {f6}
{\includegraphics [width=10cm] {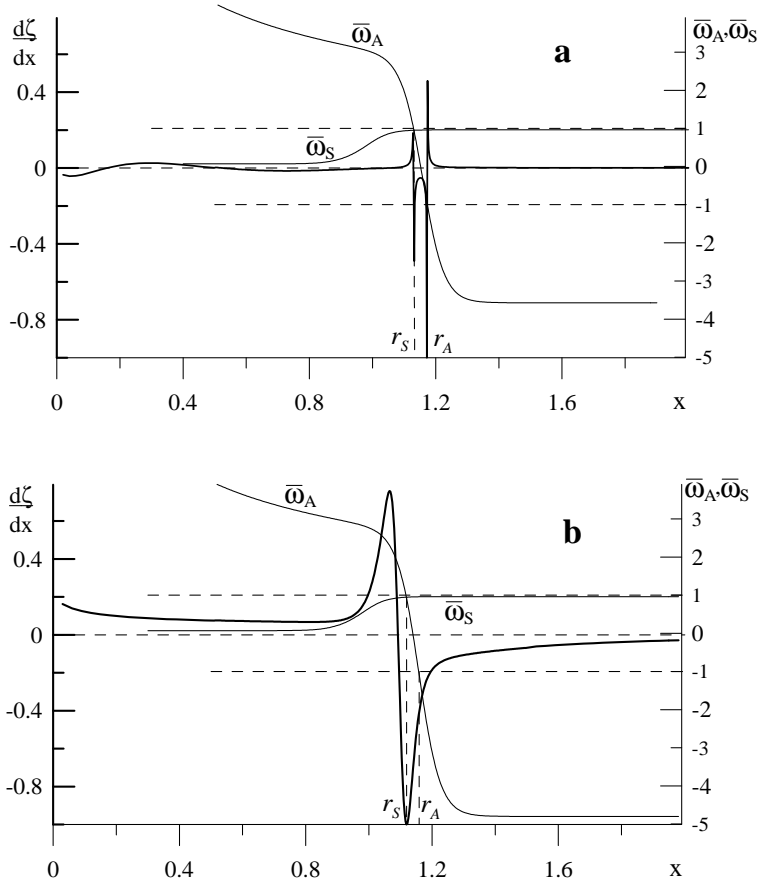}}
\caption {Spatial structure of unstable oscillations of azimuthal
harmonic $m=1$ normalised to the maximum $ |d\zeta/d x | _ {max}
$: (a) oscillations close to the second harmonic $n=2$ of eigen-modes
propagating in the FMS-waveguide in the plasma jet ($k_zr_b=2$), (b)
oscillations of the "global mode" for rather small magnitudes of
$k_zr_b\rightarrow 0$. Resonance surfaces for the Alfven and SMS oscillations
are determined by the equations $ \bar {\omega} _A (r_A) = \pm 1$ and $ \bar
{\omega} _A (r_S) = \pm \bar {\omega} _s (r_S) $ (the signs $ \pm $
corresponds to the signs of $k_z $).
}
\end {figure}

Thin lines in fig.7 depict the dependences of the MHD-oscillation increment
due to local instability of the plasma jet boundary for the first azimuthal
harmonics $m=0,1,2,3$ and for several values of the parallel wave number
$k_zr_b $. Each of the four panels displays the distribution of the
oscillation increment of the same magnitude $k_zr_b=2$ as in figs. 5 and 6.
Comparing the increments of various azimuthal harmonics, one's attention is
drawn to the fact that when $m $ increases the absolute magnitude of $
{\rm Im} (c) $ and the size of the first region of existence of unstable
oscillations $1 <M_A <M_2$ also increases. This is interpreted in the following
way, when $m $ increases the $M_2$ point is displaced into the region of large
values of $M_A $. The distributions of $ {\rm Im} (c (M_A)) $ presented in
the same panel for oscillations with $k_zr_b <1$ show that the same
happens also when the $k_zr_b $ parameter decreases \--- the $M_2$ point
shifts to the region of large magnitudes of $M_A $ (except the harmonic $m=0$
for which $M_2=1$). This is completely consistent qualitatively with what has been
obtained in the WKB approximation in the previous Section.

\section {Instability of global modes of the plasma jet oscillations}

As follows from fig.7, there is yet another type of unstable oscillations of
the plasma jet, with the increment growing when $M_A\rightarrow 0$.
The corresponding distributions of $ {\rm Im} (c (M_A)) $ are presented in
fig.7 in bold lines. The following features of these oscillations attract
one's attention:
\begin {enumerate}
  
 \item For oscillations with $m\ne 0$, instability only takes place
for rather small magnitudes of $k_zr_b <(k_zr_b) _ {max} <1$ (the magnitude
of $ (k_zr_b) _ {max} $ is different for various azimuthal harmonics $m
$).

\item The $ {\rm Im} (c (M_A)) $ plots for $m\ne 0$ practically coincide
for any unstable oscillations with $k_zr_b <(k_zr_b) _ {max} $.

\item When $M_A $ increases, the increment of the oscillations decreases
and for $m\ne 0$ it tends to zero when $M_A=M _ {Ac} $ (the magnitude of $M _
{Ac} $ is different for the different azimuthal harmonics $m $).

\item For oscillations with $m=0$ the $ {\rm Im} (c (M_A)) $ plots
differ essentially for oscillations with various $k_zr_b $ , having
no restricting value of $M _ {Ac} $ to terminate the region of existence of
unstable oscillations.

\item The absolute values of the increment are much larger for
oscillations with $m\ne 0$ than for those with $m=0$.

\end {enumerate}
\begin {figure} \label {f7}
\vspace {1cm}
{\includegraphics [width=18cm] {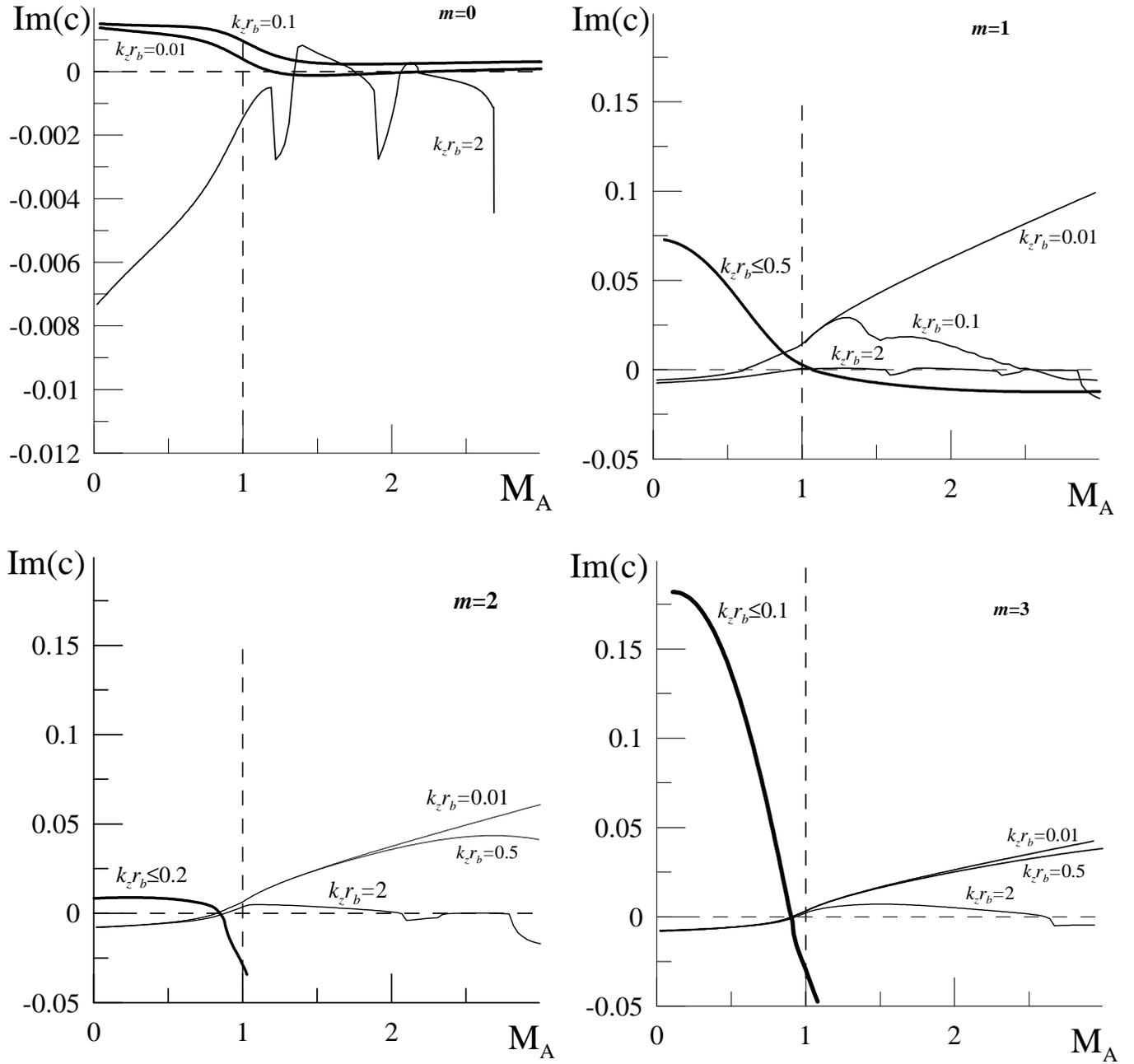}}
\caption {Mach number $M_A $ dependence of the increment of
MHD oscillations of the cylindrical plasma jet for the first azimuthal
harmonics $m=0,1,2,3$ and different magnitudes of parameter $k_zr_b $.
The distributions of the increment for the oscillations with $k_zr_b=2$ (thin
lines) defined by the local instability of the jet boundary are presented on
all the panels. For the harmonics $m\ne 0$ the distributions are also
presented of the increment of local instability for the oscillations with
$k_zr_b <1$ (thin lines). Thick lines show the distributions of the
increment of the "global modes" of the plasma jet oscillations.
}
\end {figure}

Let us try to understand qualitatively the nature and features of these
oscillations by analyzing their spatial structure in fig.6b. First of all
it should be noted that these oscillations have an almost homogeneous
structure of the first derivative $d\zeta/d r\approx const=C_I $ in the
plasma jet, whence we obtain $ \zeta=C_Ir $ (we suppose $\zeta(r\rightarrow 0)\rightarrow 0$). This means that the oscillations
occupy the entire cross-section of the plasma jet and have a constant
amplitude there. We will call such oscillations "global modes" of the
plasma jet. The dispersion equation for such modes can be obtained
using the matching conditions (\ref {Eq_6}) on the plasma jet boundary. For
this purpose we will integrate equation (\ref {Eq_4}) in the interval
($0, r_b $) using an approximate expression $ \zeta=C_Ir $ for $ \zeta $:

\[
\left.\frac {\rho_0\Omega^2} {k_r^2r} \frac {\partial r\zeta} {\partial r} \right | _ {r=r_b-\varepsilon} =-\int_0 ^ {r_b} \rho_0\Omega^2\zeta dr \approx-C_I\int_0 ^ {r_b} \rho_0\Omega^2r dr
\]
and substitute the resulting expression into the matching condition
(\ref {Eq_6}) assuming the region outside of the jet to be opaque. As
a result we obtain the following dispersion equation

\[
\int_0 ^ {r_b} \rho_0\Omega^2r dr =\frac {\rho _ {0II} \Omega _ {II} ^2r_b} {\sqrt {-k^2 _ {rII} (r_b)}},
\]
or, in the dimensionless form,

\begin {equation} \label {Eq_31}
\int_0 ^ {r_b} \left (\frac {c^2} {v_A^2 (r)}-1\right) r dr =\frac {r_b\left [(c-M_A) ^2/\epsilon^2-1\right]} {b\sqrt {-k^2 _ {rII} (r_b)}},
\end {equation}
where
 \[
k^2 _ {rII} =k_z^2\left (
\frac {(c-M_A) ^4/\epsilon^4} {(c-M_A) ^2 (1 +\beta ^ * _ {II})/\/\epsilon^2-\beta ^ * _ {II}}-1-\kappa_m^2\right),
\]
$v_A (r) =A (r)/A_I $, $b=B _ {0I} ^2/B _ {0II} ^2$, $ \epsilon=A _
{II}/A_I\ll 1$, $ \kappa_m=m/k_zr_b $.

Let us apply the perturbation theory to searching for a solution to (\ref
{Eq_31}), expanding $c $ into a (\ref {Eq_24_1})-like series with $ \epsilon
$ being the small parameter. In the zeroth order of the perturbation theory,
we have $c_0=M_A $ as before. In the first order of the perturbation theory,
we obtain an equation for $c_1$

\begin {equation} \label {Eq_32}
\frac {c_1^2-1} {\sqrt {
1 +\kappa_m^2-c_1^4 / [c_1^2 (1 +\beta ^ * _ {II})-\beta ^ * _ {II}})]} = \int_0 ^ {r_b} \left (\frac {M_A^2} {v_A^2 (r)}-1\right) rdr.
 \end {equation}
 Let us consider two cases.

 \subsection {Case $m\neq 0$}

 When the parameter $k_zr_b$ is small enough ($ \kappa_m\rightarrow \infty $,
$k_zr_b\rightarrow 0$) we obtain the following approximate solution of (\ref
{Eq_32}):

\[ c_1^2\approx bm\int_0^1\left (\frac {M_A^2} {v_A^2 (x)}-1\right) xdx,
\]

where $x=r/r_b $ and $b\gg 1$. It is easy to see from here
that the unstable oscillations ($c_1^2 <0$) exist only when the condition

\[
  \int_0 ^ {r_b} \rho_0 (r) \left (v _ {0II} ^2-A^2 (r) \right) rdr <0, \]
holds, which also determines the magnitude of $M _ {Ac} $. When
$k_zr_b\rightarrow 0$ the magnitude of $ {\rm Im} (c) $ does not depend on
$k_zr_b $, which agrees completely with the numerical calculations
in the previous Section.

 \subsection {Case $m=0$}

If $ |c_1 |\gg 1$ (for example when $b\gg 1$, but $ | \epsilon c_1 |\ll c_0$)
we have the following solution of (\ref {Eq_32})

\[
  c_1\approx
  \pm\frac {i} {\sqrt {1 +\beta ^ * _ {II}}} bk_zr_b\int_0^1\left (\frac {M_A^2} {v_A^2 (x)}-1\right) xdx.
  \]
There is no limiting magnitude of $M _ {Ac} $ for this solution, defining the
region of existence for unstable oscillations, but there is a dependence on
$k_zr_b $. This also agrees qualitatively with the above numerical
calculations. Note that the solutions obtained in this way should only be
regarded as an illustration of the qualitative behaviour of the global
oscillation modes. The exact values of their increment obtained
numerically may differ considerably from these simplified estimates.

\section {Conclusion}

Let us enumerate the main results of this work.

1. Qualitative analysis of the solution of (\ref {Eq_4}), describing
the MHD oscillations of a cylindrical plasma jet, has shown that its boundary
is unstable in relation to the fast magnetosonic oscillations in the range of
the flux parameters $1 <M_A <M_2$ (where $M_A=v _ {0II}/A_I $ is the Mach
number found from the maximum Alfven speed, $M_0 <M_1 <1$; $M_2> 1$ are
the characteristic Mach numbers defined in Section 5 in the
WKB approximation over the radial coordinate).

  2. The boundary of the plasma jet also becomes unstable when $M_A $
approachs one of the eigenvalues $c _ {0n} = {\rm Re} ((\omega/k_z) _n/A_I)
$ defined by the dispersion equation for FMS and SMS oscillations in the
resonator: $ \tan (\psi (c _ {0n}) + \pi/4) =0$, $n=1,2,3... $, where $ \psi
$ is spatial oscillation phase from the turning point to the plasma jet
boundary, $ (\omega/k_z) _n $ is the parallel phase velocity of the $n $\--th
harmonic of the oscillations. The ranges of $M_A $ where the
unstable oscillations exist are located close to $c _ {0n} $ -
in the range $M_A> M_2> 1$ for FMS waves, and $M_0 <M_A <M_1
<1$ for SMS-waves.

3. A numerical solution of the problem for a cylindrical plasma jet with
a smooth boundary layer has shown that the range of $M_A $ values where the
jet boundary is unstable corresponds qualitatively to those obtained
in the WKB approximation for the jet with a sharp boundary. There are
essential differences, however. Firstly, the absolute values of the increment
of unstable oscillations for the jet with a smooth boundary layer are
much smaller than those for the jet with a sharp boundary. This is explained
by a smoothing of the boundary and by competition between the instability of
the shear flow and the dissipation of the oscillation energy on the resonance
surfaces for the Alfven and SMS waves. The number of unstable harmonics of the
eigen-oscillations of magnetosonic resonators in the plasma jet is restricted
by a few first harmonics. Secondly, the ranges of the $M_A $ values for
unstable modes of the oscillations in the jet with a smooth boundary are
considerably displaced relative to the location obtained for
the jet with a sharp boundary.

4. It is shown that, in addition to local instability of the jet boundary,
there exist also unstable "global modes" of the plasma jet oscillations.
Their amplitude is almost constant in the entire cross-section of the jet,
while its instability increments are much larger than those for unstable
oscillations of the jet boundary. The range of $M_A $ values for which the
"global modes" are unstable begins from $M_A=0$. The distribution of the
increment of such oscillations differs substantially between the axisymmetric
($m=0$) and asymmetrical ($m\neq 0$) modes. Modes with $m\neq 0$ only become
unstable for rather small magnitudes of the parallel wave number
($k_zr_b <(k_zr_b) _ {max} <1$), the region of their existence being
restricted by the range $0 <M_A <M _ {Ac} $ (where $M _ {Ac} $ is the limiting
magnitude $M_A $ until which the "global modes" remain unstable, different
for each $m $). The distribution of the increment of such oscillations is almost
the same for any oscillations with $k_zr_b <(k_zr_b) _ {max} $. Unstable
axisymmetric "global modes" are not limited in
the $M_A $ value and their increment depends essentially on
the magnitude of parameter $k_zr_b $.

\section * {Acknowledgments}

This work was partially supported by grant 09-02-00082 and by Program of
presidium of Russian Aca\-demy of Sciences \#16 and OFN RAS \#2.16.

\section * {References}

\end {document}